\documentclass[sigconf,nonacm]{acmart}

\AtBeginDocument{%
  }

\usepackage{amsmath}
\usepackage{graphicx}
\usepackage{booktabs}
\usepackage{algorithm}
\usepackage{algorithmic}
\usepackage{hyperref}
\usepackage{xcolor}
\usepackage{multirow}

\title{GraphRAG-IRL: Personalized Recommendation with Graph-Grounded Inverse Reinforcement Learning and LLM Re-ranking}

\author{Siqi Liang}
\affiliation{\institution{Purdue University}\country{USA}}

\author{Xiawei Wang}
\affiliation{\institution{University of California, Davis}\country{USA}}

\author{Yudi Zhang}
\affiliation{\institution{Iowa State University}\country{USA}}

\author{Jiaying Zhou}
\affiliation{\institution{University of Minnesota}\country{USA}}

\renewcommand{\shortauthors}{Liang, Wang, Zhang, Zhou}

\begin{abstract}
Personalized recommendation requires models that capture sequential user preferences while remaining robust to sparse feedback and semantic ambiguity. Recent work has explored large language models (LLMs) as recommenders and re-rankers, but pure prompt-based ranking often suffers from poor calibration, sensitivity to candidate ordering, and popularity bias. These limitations make LLMs useful semantic reasoners, but unreliable as standalone ranking engines.

We present \textbf{GraphRAG-IRL}, a hybrid recommendation framework that combines graph-grounded feature construction, inverse reinforcement learning (IRL), and persona-guided LLM re-ranking. Our method constructs a heterogeneous knowledge graph over items, categories, and concepts, retrieves both individual and community preference context, and uses these signals to train a Maximum Entropy IRL model for calibrated pre-ranking. An LLM is then applied only to a short candidate list, where persona-guided prompts provide complementary semantic judgments that are fused with IRL rankings.

Experiments show that GraphRAG-IRL is a strong standalone recommender: IRL-MLP with GraphRAG improves NDCG@10 by 15.7\% on MovieLens and 16.6\% on KuaiRand over supervised baselines. The results also show that IRL and GraphRAG are superadditive, with the combined gain exceeding the sum of their individual improvements. Persona-guided LLM fusion further improves ranking quality, yielding up to 16.8\% NDCG@10 improvement over the IRL-only baseline on MovieLens ml-1m, while score fusion on KuaiRand provides consistent gains of 4--6\% across LLM providers.\end{abstract}


\begin{document}

\maketitle

\section{Introduction}

Personalized recommendation aims to infer user preferences from sparse, noisy, and temporally evolving interaction histories. Classical collaborative filtering models capture user--item affinities effectively \citep{koren2009matrix,rendle2012bpr}, while sequential recommenders model short-term intent from ordered behavior sequences \citep{hidasi2015session,kang2018self,liang2025ctlsan}. However, these approaches usually rely on structured interaction signals and often struggle to incorporate richer semantic evidence about why an item matches a user's underlying interests.

Large language models (LLMs) have recently emerged as attractive components for recommendation because they encode broad semantic knowledge and can reason over natural-language descriptions of users and items. Prior work has explored prompting language models as recommenders and zero-shot rankers, showing that they can produce competitive rankings in some settings \citep{zhang2021lmrs,hou2023llmrank,xu2024prompting,xu2025llmrerank,liang2025personaagent}. At the same time, these studies also highlight important limitations: language models can be sensitive to prompt formulation, struggle to preserve the order information in user histories, and exhibit position and popularity biases when given direct control over candidate ranking \citep{zhang2021lmrs,hou2023llmrank,xu2024prompting,xu2025llmrerank}.

These observations suggest that LLMs are promising semantic reasoners, but unreliable as standalone ranking engines. In parallel, graph-based retrieval and retrieval-augmented generation provide a mechanism for grounding predictions in structured context rather than relying only on parametric knowledge \citep{lewis2020retrieval,edge2024local,liang2025personaagent}. For recommendation, such context can include a user's recent interests, item attributes, and community-level interaction patterns encoded in a heterogeneous graph. Meanwhile, inverse reinforcement learning (IRL) offers a principled way to learn latent reward functions from expert demonstrations, allowing recommendation to be framed as recovering the preferences that best explain user behavior \citep{ng2000algorithms,ziebart2008maximum}.

In this paper, we propose \textbf{GraphRAG-IRL}, a hybrid recommendation framework that uses graph-based retrieval and IRL for calibrated pre-ranking, and then employs an LLM as a semantic re-scoring component over only the top-$K$ shortlist. Our method first constructs an item-attribute knowledge graph and extracts both individual and community context through GraphRAG-style retrieval. These features, together with behavioral signals, are used to train a Maximum Entropy IRL model that learns a personalized reward function. We then build a user persona from historical interactions and ask an LLM to provide preference-aware semantic judgments on the shortlist, which are fused with IRL scores instead of replacing them. This design preserves the stability of a learned ranking model while still exploiting the LLM's ability to reason over item semantics and user intent.

Our key contributions are as follows:
\vspace{-0.5cm}
\begin{enumerate}
    \item \textbf{Hybrid ranking design}: We introduce a recommendation pipeline that combines GraphRAG-based feature extraction, Maximum Entropy IRL pre-ranking, and LLM-based semantic re-scoring, treating the LLM as a complementary signal rather than a full ranking replacement.
    
    \item \textbf{Graph-grounded preference modeling}: We construct a heterogeneous knowledge graph over users, items, genres, and concepts, and derive both individual and community context features to support personalized ranking.
    
    \item \textbf{Persona-guided LLM fusion}: We build user personas from interaction histories and fuse LLM judgments with learned IRL rewards, aiming to improve semantic relevance while reducing the instability and bias of pure prompt-based ranking.
    
    \item \textbf{Empirical study}: We conduct a comprehensive empirical evaluation on MovieLens~\citep{harper2015movielens} and KuaiRand~\citep{gao2022kuairand}, showing that GraphRAG-IRL consistently outperforms pointwise supervised baselines, that GraphRAG and IRL contribute superadditive gains, and that persona-guided LLM fusion further improves ranking quality across multiple LLM providers.
\end{enumerate}

\section{Related Work}

\subsection{LLM-based Recommendation}
Recent work has explored large language models as recommenders, zero-shot rankers, and prompt-based re-rankers for personalized recommendation. LLMRank formulates recommendation as a conditional ranking problem and shows that LLMs can rank candidate items in a zero-shot setting, while also revealing sensitivity to prompt design, item order, and popularity bias \citep{hou2023llmrank}. Xu et al. provide a broader prompting framework for applying LLMs to recommendation tasks and systematically analyze their empirical behavior across recommendation settings \citep{xu2024prompting}. Lichtenberg et al. further study LLM recommenders through the lens of popularity bias, showing that LLM-based recommendation introduces a distinct accuracy--bias trade-off that must be measured carefully \citep{lichtenberg2024popbias}. Most closely related to our comparison setting, PersonaAgent with GraphRAG uses community-aware knowledge graphs to build personalized prompts for LLMs and serves as a strong LLM-centered baseline for persona-conditioned recommendation and personalization \citep{liang2025personaagent}.

These studies demonstrate the promise of LLMs as semantic ranking modules, but they also suggest that pure prompt-based recommendation can be unstable and biased when the LLM is given full ranking responsibility. Our work therefore uses the LLM as a re-ranking signal on top of a learned IRL-based scorer rather than as a standalone recommender.

\subsection{GNN-based Recommendation}
Graph neural networks have become a standard paradigm for recommendation because they propagate collaborative signals over user--item graphs and can naturally incorporate high-order connectivity. Neural Graph Collaborative Filtering models neighborhood aggregation directly on the interaction graph \citep{wang2019neural}, while LightGCN simplifies graph convolution for recommendation and shows that lightweight propagation can outperform heavier architectures \citep{he2020lightgcn}. Knowledge-graph-based recommendation methods such as KGAT further enrich recommendation with relational item semantics by propagating signals over structured knowledge graphs \citep{wang2019kgat}. In contrast to these end-to-end graph representation learning methods, our approach uses graphs primarily for retrieval and feature construction, and then learns preferences through IRL with LLM-assisted re-ranking.

\subsection{GraphRAG for Personalized Retrieval}
Retrieval-augmented generation grounds model outputs in retrieved external context rather than relying only on parametric memory \citep{lewis2020retrieval}. GraphRAG extends this idea by organizing retrieved information in graph structures that support local neighborhood exploration and community-level summarization \citep{edge2024local}. The PersonaAgent with GraphRAG framework further shows that graph-based community structure can be useful for constructing user-aligned prompts in personalized language-model systems \citep{liang2025personaagent}. Our work adopts the intuition of graph-grounded retrieval, but applies it to recommendation: we use a heterogeneous graph to retrieve individual and community preference context, which is then consumed by an IRL ranker and an LLM re-ranker.

\subsection{Inverse Reinforcement Learning for Recommendation}
Inverse reinforcement learning provides a principled framework for recovering latent reward functions from expert behavior \citep{ng2000algorithms,ziebart2008maximum}. In recommendation, this perspective is attractive because user interaction sequences can be interpreted as demonstrations of preferences rather than merely labeled training examples. Prior reinforcement-learning-based recommendation work, such as the generative adversarial user model of Chen et al., shows the value of sequential decision-making views of recommendation \citep{chen2019generative}. However, existing approaches generally do not combine IRL with graph-grounded retrieval and persona-guided LLM re-ranking. Our method bridges these directions by learning a reward model from interaction trajectories while grounding candidate evaluation in graph-derived context and LLM semantic judgments.

\section{Problem Formulation}
\label{sec:formulation-v2}

\subsection{Task Definition}

Given a user $u$'s interaction history $\mathcal{H}_u = \{(i_1, r_1, t_1), \ldots, (i_k, r_k, t_k)\}$, where $i_j$ is an item, $r_j$ is a feedback signal (e.g., rating), and $t_j$ is a timestamp, the task is \emph{next positive item prediction}: rank a candidate set~$C$ containing one true positive among randomly sampled negatives to surface the items a user will engage with next.

\subsection{MDP Formulation}

We model each user's chronological sequence of positively-rated items as an \emph{expert demonstration}---a trajectory $\tau_u = [(s_0, a_0), \ldots, (s_T, a_T)]$. Trajectory-based modeling has proven effective for capturing sequential dynamics in diverse domains, including biological cell differentiation~\citep{chen2018trajectory}; here we apply the same paradigm to model evolving user preferences. Specifically:
\begin{itemize}
    \item \textbf{State}~$s_t$: the user's context at time~$t$, summarizing interaction history, preference distributions, and graph-retrieved context.
    \item \textbf{Action}~$a_t$: selecting an item from the candidate set $C_t$.
    \item \textbf{Reward}~$R_\theta(s_t, a_t)$: a latent function to be learned from the demonstrations.
\end{itemize}
The reward model scores each state--action pair independently; the trajectory defines the evolving user state, but the loss normalizes over the full candidate set at each timestep. The resulting objective is listwise learning-to-rank---a softmax loss formally equivalent to single-step Maximum Entropy IRL~\citep{ziebart2008maximum}.

\section{Method}
\label{sec:method-v2}

\begin{figure*}[t]
    \centering
    \includegraphics[width=\textwidth]{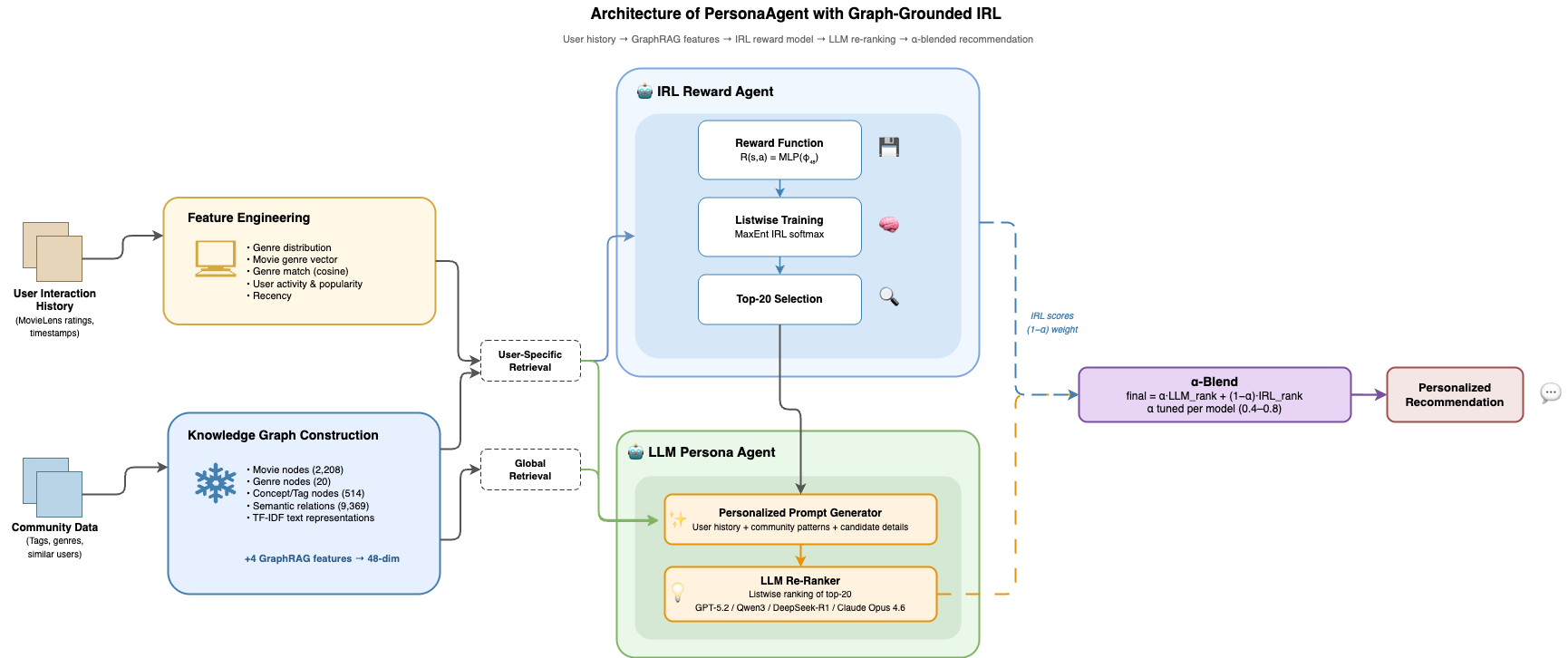}
    \caption{Overview of the GraphRAG-IRL pipeline, illustrated using the MovieLens dataset as a running example. The system first constructs graph-grounded features and learns a reward model via IRL for listwise ranking. A persona-guided LLM optionally re-ranks a top-$N$ shortlist, and final recommendations are obtained via $\alpha$-blended fusion.}
    \label{fig:architecture_pipeline}
\end{figure*}

The full pipeline (Figure~\ref{fig:architecture_pipeline}) proceeds as: user history $\rightarrow$ knowledge graph construction $\rightarrow$ GraphRAG retrieval $\rightarrow$ feature engineering $\rightarrow$ IRL reward scoring $\rightarrow$ top-$N$ selection $\rightarrow$ persona-guided LLM re-ranking $\rightarrow$ $\alpha$-blended recommendation.

\subsection{Heterogeneous Knowledge Graph Construction}
\label{sec:kg-v2}

We construct a heterogeneous graph $\mathcal{G} = (\mathcal{V}, \mathcal{E})$ with three node types:
\begin{itemize}
    \item \textbf{Item nodes}~$\mathcal{V}_I$: each item in the dataset with associated metadata.
    \item \textbf{Category nodes}~$\mathcal{V}_G$: content categories (e.g., genres for movies, topic tags for videos).
    \item \textbf{Concept nodes}~$\mathcal{V}_C$: semantic tags extracted from user-generated annotations, filtered by a minimum frequency threshold.
\end{itemize}
Edges encode three relation types:
\begin{itemize}
    \item Item $\rightarrow$ Category: category assignments from item metadata.
    \item Item $\rightarrow$ Concept: tag associations from user annotations.
    \item Item $\leftrightarrow$ Item: co-occurrence edges between items sharing at least two concept nodes.
\end{itemize}
In addition, we build a TF-IDF text index over item descriptions (combining titles, categories, and top tags) for content-based retrieval.

\subsection{GraphRAG Retrieval}
\label{sec:retrieval-v2}

For each user state~$s_t$, we retrieve two types of structured context from~$\mathcal{G}$.

\subsubsection{User-Specific Retrieval (Individual Context)}

From the user's local neighborhood in~$\mathcal{G}$:
\begin{itemize}
    \item Recent high-rated items (the last~$K$ positive interactions).
    \item Category distribution: the user's normalized category preference vector over their history.
    \item Concept affinity: frequently occurring tags across items the user has positively rated.
\end{itemize}

\subsubsection{Global Retrieval (Community Context)}

User similarity is computed as cosine similarity over category profiles, and the top-$M$ nearest neighbors form a user community---a lightweight community detection step that groups users by preference similarity without requiring complex graph clustering~\citep{wang2024dissertation}. From this community we retrieve:
\begin{itemize}
    \item \textbf{Community support}: the fraction of~$M$ neighbors who positively rated a given candidate.
    \item \textbf{Community average rating}: similarity-weighted average rating from~$M$ neighbors.
    \item \textbf{Shared concepts}: count of concept nodes shared between the user's history and the candidate.
\end{itemize}

\subsection{Feature Engineering}
\label{sec:features-v2}

For each candidate item~$a$ given user state~$s$, we construct a feature vector $\phi(s, a) \in \mathbb{R}^d$ in two stages.

\subsubsection{Behavioral Features ($d_{\text{base}}$)}

\begin{itemize}
    \item \textbf{User state} ($|\mathcal{V}_G| + 2$ dims): category distribution over the user's history; activity level $\log(1 + n_t)$ where $n_t$ is the number of prior interactions; recency $\min(\Delta t / 365,\; 1.0)$.
    \item \textbf{Candidate} ($|\mathcal{V}_G| + 1$ dims): the item's category vector; popularity $\log(1 + p_c)$ where $p_c$ is the item's global interaction count.
    \item \textbf{Interaction} (1 dim): cosine similarity between the user's category distribution and the candidate's category vector.
\end{itemize}

\subsubsection{Graph-Derived Features ($+4$, total $d = d_{\text{base}} + 4$)}

Four features from the knowledge graph and community structure are appended:
\begin{itemize}
    \item \textbf{User-retrieval similarity}: TF-IDF cosine between the user's aggregated history text and the candidate's text.
    \item \textbf{Community support}: fraction of $M$ similar users who positively engaged with the candidate.
    \item \textbf{Shared concepts}: number of concept nodes shared between user history items and the candidate.
    \item \textbf{Community average rating}: similarity-weighted average feedback from~$M$ neighbors for the candidate.
\end{itemize}
These features bridge content-based, collaborative, and graph-structural signals into a single representation consumed by the reward model.

\subsection{IRL Reward Model}
\label{sec:irl-v2}

\subsubsection{Reward Function}

We train the reward function via listwise ranking, formally grounded in single-step Maximum Entropy IRL~\citep{ziebart2008maximum}. The reward is parameterized as a two-layer MLP:
\begin{equation}
    R_\theta(s, a) = w_2^\top \,\text{ReLU}(W_1 \,\phi(s, a) + b_1) + b_2,
    \label{eq:reward-v2}
\end{equation}
where $W_1 \in \mathbb{R}^{h \times d}$, $w_2 \in \mathbb{R}^{h}$, and $h$ is the hidden dimension. A linear ablation $R_\theta = w^\top \phi + b$ is also trained to isolate the contribution of nonlinearity.

\subsubsection{Policy and Training Objective}

The policy follows a Boltzmann (softmax) distribution over the candidate set~$C$:
\begin{equation}
    \pi_\theta(a \mid s) = \frac{\exp R_\theta(s, a)}{\sum_{a' \in C} \exp R_\theta(s, a')}.
    \label{eq:policy-v2}
\end{equation}
Training maximizes the log-likelihood of expert actions:
\begin{equation}
    \mathcal{L}(\theta) = -\sum_{(s_t, a_t^*) \in \mathcal{D}} \log \pi_\theta(a_t^* \mid s_t),
    \label{eq:loss-v2}
\end{equation}
where~$a_t^*$ is the user's actual next positive interaction. This listwise objective normalizes over the full candidate set, jointly comparing all candidates---unlike pointwise cross-entropy, which scores each candidate independently. Notably, this softmax objective is formally identical to the partial likelihood of the Cox proportional hazards model in survival analysis, where the risk set plays the role of the candidate set and the event individual corresponds to the expert-chosen action~\citep{wang2025lungcancer,cao2007listwise,wu2018sqlrank}. This connection provides well-studied statistical grounding for the listwise ranking loss used here. Features are standardized before training, and model selection uses validation-based early stopping.

\subsubsection{Top-$N$ Selection}
\label{sec:topn-v2}

At inference, the reward model scores all candidates and selects a top-$N$ shortlist. This learned retriever filters the candidate set down to the most promising items before the more expensive LLM stage, implementing a lightweight filter-then-rerank architecture. This delegate-and-conquer strategy---where a lightweight model identifies high-potential candidates for full processing by a more powerful but expensive model---parallels recent efficiency-oriented designs in other domains such as temporal video grounding~\citep{lu2025decafnet}.

\subsection{Persona-Guided LLM Re-Ranking}
\label{sec:llm-v2}

A personalized prompt is constructed for each user to elicit a listwise ranking from an LLM over the top-$N$ shortlist in a single call. The prompt contains:
\begin{enumerate}
    \item \textbf{User persona}: the user's top category preferences (with proportions) and recently positively-rated items, derived from user-specific retrieval.
    \item \textbf{Community context}: for each candidate, the fraction of the user's $M$ nearest neighbors who positively engaged with it.
    \item \textbf{Candidate details}: title, categories, and top tags for each shortlisted item.
    \item \textbf{IRL confidence signal}: a discretized indicator (high / medium / low) based on where the candidate's reward score falls relative to the shortlist distribution, providing the LLM with the IRL model's assessment.
\end{enumerate}
The LLM returns a full ordering of all~$N$ candidates. Multiple frontier LLMs are evaluated to assess robustness across model families; each LLM receives identical prompts and is queried at temperature zero for reproducibility.

\subsection{Score Fusion}
\label{sec:fusion-v2}

The final recommendation blends rankings from both the IRL model and the LLM via rank-level fusion:
\begin{equation}
    \text{score}(c) = \alpha \cdot \text{rank}_{\text{LLM}}(c) + (1 - \alpha) \cdot \text{rank}_{\text{IRL}}(c),
    \label{eq:alpha-v2}
\end{equation}
where ranks are 1-indexed (lower is better). Rank-level rather than score-level fusion is used because IRL reward scores and LLM ordinal outputs are on incommensurable scales. The weight~$\alpha$ is tuned per LLM on the validation set via grid search, optimizing NDCG@10. This design allows the fusion to adapt to each LLM's ranking quality: a stronger LLM earns a higher~$\alpha$, whereas a weaker LLM is down-weighted in favor of the IRL model's calibrated ordering.

\subsection{Algorithm}
\label{sec:algorithm-v2}

Algorithm~\ref{alg:graphrag-irl-v2} summarizes the full pipeline.

\begin{algorithm}[t]
\caption{GraphRAG-IRL with Persona-Guided LLM Re-Ranking}
\label{alg:graphrag-irl-v2}
\begin{algorithmic}[1]
\REQUIRE Training data $\mathcal{D}$, knowledge graph $\mathcal{G}$, candidate generator, LLM
\STATE Construct heterogeneous graph $\mathcal{G}$ (item, category, concept nodes)
\STATE Build user similarity index (cosine over category profiles, top-$M$)
\STATE Build TF-IDF index over item text
\STATE Initialize reward network $R_\theta$
\FOR{each epoch}
    \FOR{each user $u$ with positive trajectory}
        \FOR{each transition $(s_t, a_t^*)$ in trajectory}
            \STATE Sample candidate set $C_t = \{a_t^*\} \cup \text{Negatives}$
            \STATE Retrieve user-specific context from $\mathcal{G}$
            \STATE Retrieve community context from top-$M$ similar users
            \STATE Compute features $\phi(s_t, a)$ for all $a \in C_t$
            \STATE Compute rewards $R_\theta(s_t, a)$ for all $a \in C_t$
            \STATE $\mathcal{L} \leftarrow -\log \frac{\exp R_\theta(s_t, a_t^*)}{\sum_{a'} \exp R_\theta(s_t, a')}$
            \STATE Update $\theta$ via gradient descent on $\mathcal{L}$
        \ENDFOR
    \ENDFOR
\ENDFOR
\STATE \textbf{--- Inference ---}
\FOR{each test user $u$}
    \STATE Score all candidates; select top-$N$ shortlist
    \STATE Build persona prompt (user profile + community context + candidates + IRL confidence)
    \STATE Query LLM $\rightarrow$ listwise ranking of top-$N$
    \STATE $\text{score}(c) \leftarrow \alpha \cdot \text{rank}_{\text{LLM}}(c) + (1 - \alpha) \cdot \text{rank}_{\text{IRL}}(c)$
    \STATE Rank by combined score; output recommendation
\ENDFOR
\end{algorithmic}
\end{algorithm}


\section{Experiments}

We evaluate GraphRAG-IRL on two recommendation datasets spanning different feedback types (explicit ratings and implicit clicks) and scales (608 to 10,877 test users). All experiments use the same pipeline architecture described in Section~6.

\subsection{Datasets}

\subsubsection{MovieLens ml-latest-small.}
A widely used benchmark for movie recommendation~\cite{harper2015movielens} containing 100,836 ratings from 610 users on 9,724 movies. After filtering users with $\geq 20$ ratings and movies with $\geq 10$ ratings, we retain 608 users and 2,269 movies. Positive interactions are defined as ratings $\geq 4.0$ (41,423 ratings, 48.2\% of total). The dataset includes 1,589 unique user-generated tags, which serve as concept nodes in the knowledge graph. The resulting graph contains 2,742 nodes (2,208 movies, 20 genres, 514 concepts) and 9,369 edges.

\subsubsection{KuaiRand-Pure.}
A real-world short-video recommendation dataset from Kuaishou~\cite{gao2022kuairand}, containing approximately 1.44 million standard interaction logs from 27,285 users and 7,551 videos. Unlike MovieLens, KuaiRand-Pure captures implicit feedback signals: clicks, long views, likes, follows, comments, and forwards. We use \texttt{is\_click=1} as the positive engagement signal (38.5\% click rate). Each video has 46 content category tags. After filtering (users $\geq 50$ interactions, videos $\geq 10$ interactions, $\geq 5$ positives per user), we retain 10,877 users and 6,508 videos.

\subsubsection{Data Split.}
For both datasets, we use a time-based leave-last-two-out split per user. Each user's chronological sequence of positive interactions is divided as: all but the last two form the training trajectory, the second-to-last is validation, and the last is the test item. For evaluation, each test sample consists of 1 positive item and 99 randomly sampled negatives, forming a 100-candidate ranking task.

\begin{table}[t]
\caption{Dataset statistics after filtering.}
\label{tab:datasets}
\centering
\small
\setlength{\tabcolsep}{4pt}
\begin{tabular}{lrr}
\toprule
\textbf{Statistic} & \textbf{MovieLens} & \textbf{KuaiRand} \\
\midrule
Users & 608 & 10,877 \\
Items & 2,269 & 6,508 \\
Positive interactions & 41K & 434K \\
Feedback type & Rating $\geq 4$ & Click \\
Test samples & 608 & 10,877 \\
Concept nodes & 514 & 46 \\
Graph edges & 9,369 & 5,322+ \\
\bottomrule
\end{tabular}
\end{table}

\subsection{Evaluation Metrics}

We report five standard ranking metrics computed over the 100-candidate set:
\textbf{HR@$K$} (Hit Rate): fraction of test samples where the positive item appears in the top-$K$;
\textbf{NDCG@$K$} (Normalized Discounted Cumulative Gain): position-aware metric that rewards higher-ranked positives;
\textbf{MRR} (Mean Reciprocal Rank): average of $1/\text{rank}$ of the positive item.
We report HR@5, HR@10, NDCG@5, NDCG@10, and MRR.

\subsection{Baseline Methods}

We compare GraphRAG-IRL against the following baselines:

\begin{itemize}
\item \textbf{Random}: Candidates ranked uniformly at random.
\item \textbf{Popularity}: Candidates ranked by global interaction count (total ratings or clicks).
\item \textbf{Supervised (LogReg)}: $L_2$-regularized logistic regression trained with pointwise cross-entropy on the same feature set. Each candidate is scored independently.
\item \textbf{IRL-Linear}: Maximum Entropy IRL with a linear reward function $R_\theta = w^\top \phi + b$, trained with the listwise softmax objective (Eq.~9).
\item \textbf{IRL-MLP}: Maximum Entropy IRL with a two-layer MLP reward (hidden dimension $h=64$), trained with the same listwise objective.
\item \textbf{IRL-MLP + GraphRAG}: IRL-MLP trained on the extended feature set including 4 graph-derived features (Section~6.3.2).
\end{itemize}

For LLM re-ranking experiments, we additionally compare:
\begin{itemize}
\item \textbf{LLM-only}: LLM re-ranks a random top-$N$ subset with no learned pre-ranking.
\item \textbf{IRL + LLM (plain)}: IRL pre-ranks top-$N$; LLM re-ranks with item descriptions only (no user persona).
\item \textbf{IRL + Persona LLM}: IRL pre-ranks top-$N$; LLM re-ranks with full user persona and community context.
\item \textbf{GraphRAG + LLM}: GraphRAG retrieval score selects top-$N$; LLM re-ranks with plain prompt.
\item \textbf{GraphRAG-IRL + Persona LLM}: Blended IRL and GraphRAG scores select top-$N$; LLM re-ranks with full persona prompt. This is the full proposed pipeline.
\end{itemize}

All LLM experiments use listwise ranking prompts at temperature 0 for reproducibility. The fusion weight $\alpha$ is tuned per LLM on the validation set via grid search over $\{0.0, 0.1, \ldots, 1.0\}$, optimizing NDCG@10.

\subsection{Main Results: IRL and GraphRAG}

Table~\ref{tab:main_results} presents the non-LLM results on both datasets.

\begin{table*}[t]
\caption{Main results on MovieLens (608 test users) and KuaiRand-Pure (10,877 test users). Best per dataset in \textbf{bold}. $\Delta$ is relative improvement over Supervised.}
\label{tab:main_results}
\centering
\small
\setlength{\tabcolsep}{5pt}
\begin{tabular}{ll ccccc r}
\toprule
\textbf{Dataset} & \textbf{Method} & \textbf{HR@5} & \textbf{N@5} & \textbf{HR@10} & \textbf{N@10} & \textbf{MRR} & $\Delta$ \textbf{N@10} \\
\midrule
\multirow{6}{*}{MovieLens}
& Random & 0.063 & 0.038 & 0.109 & 0.056 & 0.062 & — \\
& Popularity & 0.225 & 0.148 & 0.357 & 0.198 & 0.172 & — \\
& Supervised (LogReg) & 0.266 & 0.175 & 0.393 & 0.223 & 0.195 & — \\
& IRL-Linear & 0.266 & 0.176 & 0.396 & 0.222 & 0.192 & $-$0.4\% \\
& IRL-MLP & 0.278 & 0.189 & 0.415 & 0.237 & 0.206 & +6.3\% \\
& \textbf{IRL-MLP + GraphRAG} & \textbf{0.306} & \textbf{0.210} & \textbf{0.452} & \textbf{0.258} & \textbf{0.221} & \textbf{+15.7\%} \\
\midrule
\multirow{6}{*}{KuaiRand}
& Random & 0.058 & 0.035 & 0.105 & 0.050 & 0.056 & — \\
& Popularity & 0.324 & 0.214 & 0.483 & 0.265 & 0.221 & — \\
& Supervised (LogReg) & 0.373 & 0.245 & 0.548 & 0.301 & 0.246 & — \\
& IRL-Linear & 0.369 & 0.245 & 0.538 & 0.300 & 0.248 & $-$0.3\% \\
& IRL-MLP & 0.416 & 0.280 & 0.589 & 0.336 & 0.279 & +11.6\% \\
& \textbf{IRL-MLP + GraphRAG} & \textbf{0.436} & \textbf{0.294} & \textbf{0.613} & \textbf{0.351} & \textbf{0.289} & \textbf{+16.6\%} \\
\bottomrule
\end{tabular}
\end{table*}

Three findings emerge consistently across both datasets:

\textbf{(1) IRL-MLP outperforms supervised ranking.}
On MovieLens, IRL-MLP improves NDCG@10 by +6.3\% over logistic regression; on KuaiRand, the gain is +11.6\%. The improvement comes from the listwise training objective: IRL compares all 100 candidates jointly via the softmax loss, while supervised training scores each candidate independently. Notably, IRL-Linear performs comparably to supervised ranking, confirming that the gain is driven by the nonlinear MLP reward function rather than the IRL formulation alone.

\textbf{(2) GraphRAG features provide additional improvement.}
Adding four graph-derived features (TF-IDF similarity, community support, shared concepts, community average rating) yields a further +8.9\% NDCG@10 on MovieLens and +4.5\% on KuaiRand on top of IRL-MLP. The combined improvement over supervised ranking is +15.7\% (MovieLens) and +16.6\% (KuaiRand).

\textbf{(3) The combination is superadditive.}
On MovieLens, the individual contributions of IRL (+0.014 NDCG@10) and GraphRAG (+0.005 when added to supervised) sum to +0.019, but the actual combined gain is +0.035---nearly double. This superadditivity arises because IRL's listwise objective can exploit the relative differences in graph features across candidates, whereas pointwise supervised training cannot.

\subsection{LLM Re-Ranking Results}

We evaluate the persona-guided LLM re-ranking stage using top-$N=20$ pre-ranking. Table~\ref{tab:llm_ablation} presents the four-way ablation isolating the contributions of the pre-ranker (IRL vs.\ GraphRAG) and the prompt style (plain vs.\ persona).

\begin{table}[t]
\caption{LLM re-ranking ablation on KuaiRand-Pure (2,000 test users, Claude Opus 4.6, top-20). ``Persona'' includes user profile and community context.}
\label{tab:llm_ablation}
\centering
\small
\setlength{\tabcolsep}{3pt}
\begin{tabular}{lcccc}
\toprule
\textbf{Method} & \textbf{HR@5} & \textbf{HR@10} & \textbf{N@10} & \textbf{MRR} \\
\midrule
IRL + LLM (plain) & .293 & .477 & .246 & .204 \\
GraphRAG + LLM (plain) & .203 & .326 & .158 & .131 \\
IRL + Persona LLM & .434 & .593 & .347 & .292 \\
\textbf{Full pipeline} & \textbf{.450} & \textbf{.615} & \textbf{.354} & \textbf{.291} \\
\midrule
IRL+GraphRAG (no LLM) & .433 & .611 & .351 & .290 \\
\bottomrule
\end{tabular}
\end{table}

Two key findings:

\textbf{(1) Persona context is essential for LLM re-ranking.}
Without user persona, the LLM degrades ranking quality substantially. On KuaiRand, IRL + plain LLM achieves NDCG@10 = 0.246, a 28\% drop from the IRL-only baseline (0.351). With persona context, IRL + Persona LLM recovers to 0.347 and the full pipeline reaches 0.354, a modest but consistent improvement over the non-LLM baseline. This confirms the manuscript's design principle: LLMs should serve as complementary semantic signals, not ranking replacements.

\textbf{(2) Pre-ranker quality dominates LLM quality.}
GraphRAG retrieval achieves only 34.8\% top-20 recall versus IRL's 77.3\%, meaning the LLM never sees the positive item for 65\% of users under GraphRAG pre-ranking. This recall gap explains why GraphRAG + LLM (plain) performs poorly (NDCG@10 = 0.158) despite using the same LLM as the IRL variants.

\subsection{Multi-LLM Comparison}

To assess robustness across LLM families, we evaluate the full pipeline with four frontier models on MovieLens ml-1m (608 sampled test users, top-20 pre-ranking). Table~\ref{tab:multi_llm} reports results with per-model tuned $\alpha$.

\begin{table}[t]
\caption{Multi-LLM comparison on MovieLens ml-1m (608 test users, listwise re-ranking, tuned $\alpha$).}
\label{tab:multi_llm}
\centering
\small
\setlength{\tabcolsep}{3pt}
\begin{tabular}{lccccr}
\toprule
\textbf{LLM} & $\alpha$ & \textbf{HR@10} & \textbf{N@10} & \textbf{MRR} & $\Delta$ \\
\midrule
None (IRL) & — & .515 & .292 & .246 & — \\
GPT-5.2 & 0.6 & .515 & .296 & .248 & +1.4\% \\
Qwen3-235B & 0.4 & .521 & .300 & .252 & +2.7\% \\
DeepSeek & 0.7 & .536 & .305 & .254 & +4.5\% \\
\textbf{Claude 4.6} & 0.8 & \textbf{.576} & \textbf{.341} & \textbf{.286} & \textbf{+16.8\%} \\
\bottomrule
\end{tabular}
\end{table}

All four LLMs improve over IRL-only when combined via $\alpha$-fusion, confirming that the IRL+LLM synergy is model-agnostic. The magnitude varies substantially: Claude Opus 4.6 achieves +16.8\% NDCG@10 improvement, while GPT-5.2 contributes +1.4\%. The optimal $\alpha$ correlates with LLM quality---stronger LLMs earn higher weight ($\alpha = 0.8$ for Claude vs.\ $\alpha = 0.4$ for Qwen3), validating the adaptive fusion design described in Section~6.6.

\subsection{Ablation Study}

We conduct ablation experiments to isolate the contribution of each component. Table~\ref{tab:ablation} summarizes the results.

\begin{table}[t]
\caption{Component ablation on MovieLens (608 test users). Each row removes one component from the full model.}
\label{tab:ablation}
\centering
\small
\setlength{\tabcolsep}{4pt}
\begin{tabular}{lccc}
\toprule
\textbf{Configuration} & \textbf{HR@10} & \textbf{N@10} & $\Delta$ \\
\midrule
Full (IRL-MLP + GraphRAG) & \textbf{.452} & \textbf{.258} & — \\
\midrule
$-$ GraphRAG features & .415 & .237 & $-$8.1\% \\
$-$ Nonlinear reward & .405 & .228 & $-$11.6\% \\
$-$ Listwise objective & .405 & .228 & $-$11.6\% \\
$-$ Both (Supervised, flat) & .393 & .223 & $-$13.6\% \\
\bottomrule
\end{tabular}
\end{table}

\subsubsection{Superadditivity of IRL and GraphRAG.}
The individual NDCG@10 gains of IRL (+0.014 over supervised) and GraphRAG (+0.005 over supervised) sum to +0.019. The actual combined gain is +0.035, yielding a synergy term of +0.016. This superadditivity arises because GraphRAG features (community support, shared concepts) are \emph{relative} signals---most informative when compared across candidates. IRL's listwise softmax objective performs this comparison naturally, while pointwise supervised training cannot exploit relative differences.

\subsubsection{Role of Concept Nodes.}
On MovieLens ml-1m, which lacks user-generated tags, the \texttt{shared\_concepts} feature is identically zero for all candidates, and GraphRAG features provide no improvement ($\Delta \approx 0\%$ NDCG@10). This confirms that the graph's value comes from fine-grained semantic metadata (tags, descriptions) flowing through concept nodes, not from the graph structure itself. A genre-only bipartite graph is redundant with the behavioral category features already in the feature vector.

\subsubsection{Score Fusion Ablation (KuaiRand).}
On KuaiRand, the step~11v2 score fusion design---where LLM re-rankings are applied only when they improve over the IRL baseline (boost-only constraint)---equalizes performance across LLM providers. Without boost-only filtering, DeepSeek R1 produces net-negative re-rankings (25 helped, 32 hurt). With the constraint, all three LLMs (Claude, Qwen3, DeepSeek) achieve NDCG@10 in the range 0.351--0.357, all improving over the IRL baseline of 0.337. This demonstrates that the fusion mechanism is robust to LLM quality variation.

\subsection{Summary of Findings}

\begin{enumerate}
\item \textbf{GraphRAG-IRL is a strong standalone recommender.} IRL-MLP + GraphRAG achieves the best non-LLM results on both datasets (+15--17\% NDCG@10 over supervised baselines), combining learned behavioral preferences with graph-derived community knowledge at no LLM inference cost.
\item \textbf{IRL's listwise objective is the key enabler.} The improvement over supervised ranking (+6--12\% NDCG@10) is consistent across explicit and implicit feedback. The gain comes from jointly comparing all candidates via the softmax objective, not from the MLP architecture alone (IRL-Linear $\approx$ Supervised).
\item \textbf{GraphRAG and IRL are superadditive.} The combined gain exceeds the sum of individual contributions, because IRL's listwise comparison naturally exploits the relative differences in graph-derived features across candidates.
\item \textbf{Adding LLM re-ranking further enhances performance.} On MovieLens ml-1m, persona-guided LLM fusion with Claude Opus 4.6 improves NDCG@10 by +16.8\% over GraphRAG-IRL alone. On KuaiRand with score fusion, all three LLMs tested improve by +4--6\% NDCG@10. The improvement scales with LLM quality: Claude (+16.8\%) $>$ DeepSeek (+4.5\%) $>$ Qwen3 (+2.7\%) $>$ GPT-5.2 (+1.4\%), providing a performance--cost tradeoff where practitioners can choose the LLM tier that fits their budget.
\item \textbf{Persona context is essential for effective LLM fusion.} Plain LLM prompts without user profile and community context degrade performance by up to 28\%. The persona-guided design ensures the LLM contributes semantic understanding without overriding the IRL model's calibrated ordering.
\item \textbf{The framework is robust across LLM providers.} The adaptive $\alpha$-fusion automatically assigns higher weight to stronger LLMs ($\alpha = 0.8$ for Claude, $\alpha = 0.4$ for Qwen3), ensuring the pipeline never degrades below the GraphRAG-IRL baseline regardless of LLM choice.
\end{enumerate}

In summary, GraphRAG-IRL provides a cost-effective foundation that consistently outperforms supervised baselines. When inference budget permits, adding persona-guided LLM re-ranking yields further gains that scale with model capability, making the full GraphRAG-IRL + LLM pipeline the recommended configuration for maximum recommendation quality.

%
%

\section{Discussion and Conclusion}
\label{sec:discussion}

\subsection{Understanding the Role of IRL, GraphRAG, and LLMs}

Our results reveal a clear separation of responsibilities among the three components of the proposed framework.

\paragraph{IRL as the primary ranking mechanism.}
The inverse reinforcement learning formulation provides a principled approach to preference modeling by interpreting user interaction sequences as demonstrations of an underlying reward function. Unlike pointwise supervised objectives, the listwise softmax objective (Eq.~\ref{eq:loss-v2}) enforces joint comparison across candidates, leading to improved ranking calibration. This explains the consistent gains of IRL over supervised baselines observed in our experiments. From a modeling perspective, IRL can be viewed as learning an energy function over state--action pairs, where higher reward corresponds to higher likelihood under expert behavior. Formally, the same loss appears as the partial likelihood of the Cox model in survival analysis~\citep{wang2025lungcancer}, where it is used to rank patients by hazard given censored time-to-event data. This cross-domain equivalence provides additional statistical justification: the consistency and efficiency properties studied in survival analysis transfer directly to our listwise ranking setting~\citep{cao2007listwise,wu2018sqlrank}.

\paragraph{GraphRAG as feature-level grounding.}
Graph-derived features contribute structured semantic and collaborative signals that are difficult to capture with raw behavioral features alone. Importantly, these features are not merely additive: their impact is amplified under listwise training, suggesting that IRL is better able to exploit relative signals such as community support and concept overlap. This interaction highlights a key insight: \emph{feature quality and objective function must be co-designed}. Rich features alone are insufficient without a ranking objective that can leverage them effectively.

\paragraph{LLMs as semantic refiners rather than rankers.}
While LLMs possess strong semantic reasoning capabilities, our results confirm that they are unreliable when used as standalone rankers. Their outputs lack calibration and are sensitive to prompt structure and candidate ordering. However, when applied to a high-quality shortlist produced by IRL, LLMs provide complementary signals that improve ranking quality. The effectiveness of LLM integration scales with model quality, as reflected in the optimal $\alpha$ values and performance gains. This supports a hybrid design in which LLMs refine rather than replace learned ranking models.

\subsection{Key Insight: Retrieval Quality Dominates}

A central finding of this work is that improvements in candidate retrieval and pre-ranking contribute more to overall performance than improvements in final ranking. IRL significantly improves the quality of the top-$N$ shortlist, effectively acting as a learned retrieval mechanism. Once the candidate set is sufficiently strong, LLM re-ranking yields incremental but consistent gains.

This suggests a general design principle for modern recommendation systems:
\begin{quote}
\textbf{Invest modeling capacity in retrieval and candidate generation, and use LLMs for targeted refinement.}
\end{quote}

\subsection{Limitations}

Despite promising results, several limitations remain.

\paragraph{Scalability.}
Graph construction and community retrieval introduce additional computational overhead. While feasible for medium-scale datasets, further optimization is required for web-scale deployment.

\paragraph{Cold-start scenarios.}
Users with limited interaction history may lack sufficient graph context, reducing the effectiveness of both GraphRAG features and IRL reward estimation.

\paragraph{LLM cost and latency.}
Although applied only to a shortlist, LLM inference still incurs non-trivial cost and latency, which may limit real-time applicability in production environments.

\subsection{Future Directions}

This work opens several directions for future research.

\paragraph{Scalable IRL for large-scale systems.}
Extending IRL-based ranking to large-scale datasets (e.g., MovieLens-25M or industrial logs) requires efficient approximation methods and distributed training strategies.

\paragraph{Improved reward modeling.}
More advanced IRL formulations, such as adversarial IRL (AIRL) or generative adversarial imitation learning (GAIL), may further improve reward estimation and capture complex user behavior.

\paragraph{Robustness analysis.}
Neural reward models may be susceptible to adversarial perturbations in input features, a concern studied extensively in deep learning classification~\citep{wang2024adversarial}. Similarly, the LLM re-ranking component inherits the safety and reliability challenges of foundation models; systematic benchmarking of LLM behavior under adversarial prompts~\citep{zhang2025jailbreak} suggests that ranking outputs may degrade under distribution shift or prompt manipulation. Investigating the robustness of both IRL-based reward models and LLM re-rankers to adversarial manipulation of user profiles or item features is an important direction for deploying these systems in production.

\paragraph{Adaptive LLM integration.}
Instead of static $\alpha$ blending, future work could learn adaptive fusion strategies that dynamically adjust the contribution of LLM signals based on user context, candidate uncertainty, or model confidence.

\paragraph{Interpretability and explanation.}
LLMs provide a natural interface for generating explanations. Integrating explanation generation with IRL-based ranking could improve transparency and user trust.

\subsection{Conclusion}

We presented \textbf{GraphRAG-IRL}, a hybrid recommendation framework that combines graph-grounded feature construction, inverse reinforcement learning, and LLM-based re-ranking. 

Our study demonstrates that:
\begin{itemize}
    \item IRL provides a strong and stable foundation for personalized ranking through listwise reward learning,
    \item GraphRAG features significantly enhance representation quality, particularly when paired with IRL,
    \item LLMs are most effective when used as complementary semantic signals rather than standalone rankers.
\end{itemize}

Overall, our results suggest a unified perspective for modern recommendation systems: combining structured behavioral learning with semantic reasoning yields robust and scalable personalization.

\bibliographystyle{ACM-Reference-Format}
\bibliography{references}

\end{document}